\documentclass[journal,12pt,onecolumn,draftclsnofoot]{IEEEtran}

\usepackage{amsmath}
\usepackage{amssymb}
\usepackage{graphicx}
\usepackage{algorithm}
\usepackage{algpseudocode}
\usepackage{subcaption}
\usepackage{multicol}
\usepackage{multirow}

\graphicspath{{images/}}

\ifCLASSINFOpdf

\else

\fi

\begin{document}
% Title
\title{Fast submodular maximization subject to $k$-extendible system constraints}

\author{
\IEEEauthorblockN{Teng Li\IEEEauthorrefmark{1}, ~ Hyo-Sang Shin\IEEEauthorrefmark{1}, ~ Antonios Tsourdos\IEEEauthorrefmark{1}} \\
\IEEEauthorblockA{\IEEEauthorrefmark{1}School of Aerospace Transport and Manufacturing, Cranfield University, Bedfordshire, MK43 0AL, UK}
}

\maketitle

\IEEEpeerreviewmaketitle
% end title

% Abstract
\begin{abstract}

	As the scales of data sets expand rapidly in some application scenarios, increasing efforts have been made to develop fast submodular maximization algorithms. This paper presents a currently the most efficient algorithm for maximizing general non-negative submodular objective functions subject to $k$-extendible system constraints. Combining the sampling process and the decreasing threshold strategy, our algorithm Sample Decreasing Threshold Greedy Algorithm (SDTGA) obtains an expected approximation guarantee of ($p-\epsilon$) for monotone submodular functions and of ($p(1-p)-\epsilon$) for non-monotone cases with expected computational complexity of only $O(\frac{pn}{\epsilon}\ln\frac{r}{\epsilon})$, where $r$ is the largest size of the feasible solutions, $0<p \leq \frac{1}{1+k}$ is the sampling probability and $0< \epsilon < p$. If we fix the sampling probability $p$ as $\frac{1}{1+k}$, we get the best approximation ratios for both monotone and non-monotone submodular functions which are $(\frac{1}{1+k}-\epsilon)$ and $(\frac{k}{(1+k)^2}-\epsilon)$ respectively. While the parameter $\epsilon$ exists for the trade-off between the approximation ratio and the time complexity. Therefore, our algorithm can handle larger scale of submodular maximization problems than existing algorithms.

\end{abstract}
  
 % Introduction
\section{Introduction}

	The constrained submodular maximization problem has drawn dramatic attention in various combinatorial optimization applications. The main feature of submodularity is the diminishing returns which means the marginal value of any elements decreases as more elements have already been selected \cite{Diminishing Returns 2007}. Constrained submodular maximization refers to selecting a set of elements maximizing the submodular objective function meanwhile satisfying a certain constraint or a combination of several constraints. 
	
	This paper addresses submodular maximization problems, especially subject to  $k$-extendible system constraints. Note that the $k$-extendible system constraint is a general type of constraint that has been widely studied. The concept of $k$-extendible systems was first introduced by Mestre in 2006 \cite{k-extendible 2006}. The intersection of $k$ matroids based on the same ground set is always $k$-extendible \cite{k-extendible 2006}. Many types of constraints handled in submodular maximisation fall into the $k$-extendible system constraint: cardinality constraint, partition matroid constraint and $k$-matroid constraint are good examples of the $k$-extendible system constraint.  
	
	Since finding the optimal solution of submodular maximization is NP-hard, much effort has been made to develop fast algorithms that can provide satisfying suboptimal solutions. The greedy-based algorithms have been widely used in constrained submodular maximization problems. Typical applications of general $k$-extendible system constraints involve Maximum Traveling Salesman Problem (Max-TSP) \cite{Max-TSP 2013}, Personalized Data Summarization \cite{Fantom 2016} and Movie Recommendation \cite{Fantom 2016, Greed is Good 2017}. In terms of cardinality constraint and matroid constraint which are subclasses of $k$-extendible system constraints, representative applications include, but are not limited to, Facility Localization \cite{Facility Location 2005}-\cite{Facility Location 2015}, Data Summarization \cite{Data Summarization 2010}-\cite{Data Summarization 2015} and Robotics \cite{Robotics 2015}-\cite{Aerial 3D Scanning}.

%%%%%%%%%
%%%%%%%%%
%%%%%%%%%
	The issue is that the sizes of data sets tend to increase. NP-hard problems are known to significantly suffer from ``curse of dimensionality", which implies that the complexity of the problem explodes as the problem size increases. Therefore, the trend of increasing sizes of data sets combined with the NP-hardness of the problem urges to develop more computationally efficient optimization algorithms. To this end, there have been numerous works recently carried out to develop more efficient constrained submodular maximization algorithms and many of them endeavor to increase computational efficiency even by sacrificing some of approximation ratio. We classify these works by the type of constraints and provide a summary of their developments in the following.

	$\bullet$ \textit{Cardinality constraint:} The Decreasing Threshold Greedy proposed in \cite{Threshold 2014} obtained an approximation ratio of ($1-1/e-\epsilon$) with time complexity of $O(\frac{n}{\epsilon}\ln\frac{n}{\epsilon})$ for monotone submodular maximization which is theoretically faster than the Lazy Greedy \cite{Lazy Greedy 1978}. This is the first one-pass streaming algorithm whose time complexity is independent on the size of the largest feasible solution $r$. Later, the sampling-based Stochastic Greedy proposed by Mirzasoleiman et al. \cite{Stochastic Greedy 2015} achieved an expectantly the same approximation ratio with lower time complexity of $O(n\ln\frac{1}{\epsilon})$. The Stochastic Greedy gets orders of magnitudes faster by losing only a bit of approximation ratio. Then in \cite{Comparing apples and oranges 2016}, Buchbinder et al. extended the analysis to general non-monotone submodular functions and achieved an approximation guarantee of ($1/e-\epsilon$) with complexity of $\min\{O(\frac{n}{\epsilon^2}\ln\frac{1}{\epsilon}), O(r\sqrt{\frac{n}{\epsilon}\ln\frac{r}{\epsilon}}+\frac{n}{\epsilon}\ln\frac{r}{\epsilon})\}$.
	
	$\bullet$ \textit{Matroid constraint:} The original Greedy algorithm \cite{First Submodular 1978} provides a $\frac{1}{2}$-approximation with time complexity of $O(nr)$ for monotone submodular maximization. A close variant of the Decreasing Threshold Greedy described in \cite{Comparing apples and oranges 2016} achieves a similar approximation ratio of $(\frac{1}{2}-\epsilon)$ but is faster with time complexity of $O(\frac{n}{\epsilon}\ln\frac{r}{\epsilon})$. Additionally, the multilinear extension was utilized to increase the approximation ratio and provide the approximation guarantee for non-monotone submodular functions, however, this kind of algorithms are quite time-consuming \cite{Measured Continuous Greedy 2011}. To remedy this, the idea of Decreasing Threshold was adopted to reduce the computational complexity \cite{Threshold 2014, Smoother Measured Continuous Greedy 2017}.

	$\bullet$ \textit{$k$-extendible system constraint:} The Decreasing Threshold Greedy \cite{Threshold 2014} provides a slightly worse approximation guarantee of $\frac{1}{1+k+\epsilon}$ but lower time complexity of $O(\frac{n}{\epsilon^2}\ln^2\frac{n}{\epsilon})$ than the original Greedy algorithm \cite{First Submodular 1978} for monotone submodular functions. For non-monotone submodular functions, Gupta et al. \cite{Gupta 2010} provided an approximation ratio of $\frac{k}{(k+1)(3k+3)}$ with time complexity of $O(nrk)$. Then the approximation ratio was improved to $\frac{k}{(k+1)(2k+1)}$ by an algorithm called FANTOM proposed by Mirzasoleiman et al. \cite{Fantom 2016} with the same complexity. After this, Feldman et al. \cite{Greed is Good 2017} made a significant breakthrough in terms of time complexity. The Sample Greedy they proposed achieves an approximation ratio of $\frac{k}{(k+1)^2}$ with complexity of $O(n+nr/k)$.
	
	From the previous achievements, it is clear that although gradual improvements have been made for general $k$-extendible system constraints recently, they are not as fruitful as those for cardinality constraints and matroid constraints. An immediate research question would be whether or not we can develop a fast algorithm that can also provide some trade-off between the approximation ratio and the time complexity for maximizing general non-negative submodular functions subject to $k$-extendible system constraints. 
	
	Inspired by the sampling strategy from \cite{Greed is Good 2017} and the decreasing threshold idea from \cite{Threshold 2014}, we propose an algorithm that is even faster than the Sample Greedy \cite{Greed is Good 2017}. The proposed algorithm, which is called Sample Decreasing Threshold Greedy Algorithm (SDTGA), provides an expected approximation guarantee of ($p-\epsilon$) for monotone submodular functions and of ($p(1-p)-\epsilon$) for non-monotone cases with expected time complexity of only $O(\frac{pn}{\epsilon}\ln\frac{r}{\epsilon})$, where $0<p \leq \frac{1}{1+k}$ is the sampling probability and $0 < \epsilon < p$. If we fix the sampling probability $p$ as $\frac{1}{1+k}$, we get the best approximation ratios for both monotone and non-monotone submodular functions which are $(\frac{1}{1+k}-\epsilon)$ and $(\frac{k}{(1+k)^2}-\epsilon)$ respectively. Here, $\epsilon$ acts as a parameter for trade-off between the approximation ratio and the time complexity. The theoretical performances of our algorithm and the benchmark algorithms are compared in Table \ref{performance comparison}.

\begin{table}[H]
\caption{Performances of Algorithms for Non-monotone Submodular Maximization}
\label{performance comparison}
\centering
\begin{tabular}{|l|c|l|}
\hline
\textbf{Algorithms} 		& \textbf{Approximation Ratio} 		& \textbf{Time Complexity} \\ \hline
 FANTOM \cite{Fantom 2016}	& $\frac{k}{(k+1)(2k+1)}$ 		& $O(nrk)$ \\ \hline
 Sample Greedy \cite{Greed is Good 2017}		& $\frac{k}{(1+k)^2}$ 		& $O(n+nr/k)$ \\ \hline
 SDTGA 	& $\frac{k}{(1+k)^2}-\epsilon$ 	& $O(\frac{n}{(1+k)\epsilon}\ln\frac{r}{\epsilon})$ \\ \hline
\end{tabular}
\end{table}
	
%\par ~
%\par 
%\par ~

% Preliminaries
\section{Preliminaries}

In this section, we present some necessary definitions and basic concepts related to the proposed algorithm.

\newtheorem{definition}{Definition}
\begin{definition}
\label{def:submodularity}
(Submodularity) \cite{Greed is Good 2017} A set function $f:2^{\mathcal{N}}\rightarrow\mathbb{R}$ is \textit{submodular} if, $\forall~X,Y\subseteq\mathcal{N}$,
\begin{equation*}
f(X)+f(Y)\geq f(X\cap Y)+f(X\cup Y).
\end{equation*}
where $\mathcal{N}$ is a finite set. Equivalently, $\forall~A\subseteq B \subseteq \mathcal{N}$ and $u\in\mathcal{N}\backslash B$,
\begin{equation}
\label{def:diminishing returns}
f(A\cup\{u\})-f(A)\geq f(B\cup\{u\})-f(B).
\end{equation}
\end{definition}

\begin{definition}
\label{def:marginal_gain}
(Marginal gain) \cite{Submodular function maximization}  
For a set function $f:2^{\mathcal{N}}\rightarrow\mathbb{R}$, $S \subseteq \mathcal{N}$ and $u \in \mathcal{N}$, define the \textit{marginal gain} of $f$ at $S$ with respect to $u$ as
\begin{equation*}
\Delta f(u|S):=f(S \cup \{u\})-f(S).
\end{equation*}
\end{definition}

The inequation (\ref{def:diminishing returns}) is known as the diminishing returns which is a crucial property of submodular functions: the marginal gain of a given element $u$ will never increase as more elements have already been added into the set $S$.

\begin{definition}
\label{def:monotonicity}
(Monotonicity) \cite{Submodular function maximization}  
A set function $f:2^{\mathcal{N}}\rightarrow\mathbb{R}$ is \textit{monotone} if, $\forall A \subseteq B \subseteq \mathcal{N}$, $f(A) \leq f(B)$. $f$ is \textit{non-monotone} if it is not monotone.
\end{definition}

In this paper, we only consider normalized (i.e. $f(\emptyset)=0$)  non-negative (i.e. $f(S) \geq 0$, $\forall S \subseteq \mathcal{N}$) submodular function maximization problem, because it is impossible to achieve any approximation guarantee for maximizing submodular functions that can take negative values \cite{Maximizing non-monotone submodular functions}.

\begin{definition}
\label{def:matroid}
(Matroid) \cite{Threshold 2014} 
A matroid is a pair $\mathcal{M}=(\mathcal{N},\mathcal{I})$ where $\mathcal{N}$ is a finite set and $\mathcal{I} \subseteq 2^\mathcal{N}$ is a collection of independent sets, satisfying:
\begin{itemize}
\item $\emptyset \in \mathcal{I}$ 
\item if $A \subseteq B, B \in \mathcal{I}$, then $A \in \mathcal{I}$ 
\item if $A, B \in \mathcal{I},|A|<|B|$, then $\exists~u \in B \backslash A ~\mbox{such that}  ~A \cup \{u\} \in \mathcal{I}$
\end{itemize}
\end{definition}

%\vskip 5pt
%Specifically, the matroid constraint can be separated as uniform matroid constraint and  partition matroid constraint. The uniform matroid constraint is also called cardinality constraint which is a special case of matroid constraint where any subset $S \subseteq \mathcal{N}$ satisfying $|S|\leq k$ is independent. While the partition matroid constraint means that a subset $S$ can contain at most one element from each partition.

% copy from <Greedy is Good> !!!
\begin{definition}
\label{def:extension}
(Extension) \cite{Greed is Good 2017}
If an independent set $B$ strictly contains an independent set $A$, then $B$ is called an extension of $A$.
\end{definition}

\begin{definition}
\label{def:$k$-extendible}
($k$-extendible) \cite{k-extendible 2006}
A $k$-extendible system is an independence system $(\mathcal{N}, \mathcal{I})$ that for every independent set $A \in \mathcal{I}$, an extension $B$ of $A$ and an element $u \notin A$ and $A \cup \{u\} \in \mathcal{I}$ there exists a subset $X \subseteq B \backslash A$ with $|X| \leq k$ such that $B \backslash X \cup \{u\} \in \mathcal{I}$.
\end{definition}

	Intuitively, if an element $u$ is added into an independent set $A$ of a $k$-extendible system, it requires at most $k$ other elements to be removed from $A$ in order to keep the set independent \cite{Greed is Good 2017}.

The following is an important claim that provides the mathematical foundation for the Sample Greedy  to work well in non-monotone submodular maximization.

\newtheorem{claim}{Claim}
\begin{claim}
\label{thm:main_claim}
\cite{Random Greedy} 
Let $h:2^\mathcal{N}\rightarrow\mathbb{R}_{\geq0}$ be a submodular function, and let $S$ be a random subset of $\mathcal{N}$. If each element of $S$ appears with a probability at most $p$ (not necessarily independently). Then, $\mathbb{E}[h(S)]\geq(1-p)h(\emptyset)$.
\end{claim}

% Algorithm and Analysis
\section{Algorithm and Analysis}

	This section describes SDTGA and analyzes its performances in details.

\subsection{Algorithm}

	We combine the sampling strategy from \cite{Greed is Good 2017} and the decreasing threshold idea from \cite{Threshold 2014} and make some variants to develop our algorithm. Firstly, we sample elements from the ground set $\mathcal{N}$ with probability $p$ to form a sample subset $R$. Then we run the decreasing threshold greedy algorithm on the subset $R$. The integrated statements are described in Algorithm \ref{alg: Sample Threshold}.

	Notice that, if the marginal gain of an element is less than $\frac{\epsilon}{r}d$, then we say this element is negligible. This is where the termination condition comes from. In addition, If the marginal gain of an element is already less than $\frac{\epsilon}{r}d$ during some iteration, then it will never be greater or equal to $\frac{\epsilon}{r}d$ in the following iterations because of the property of the submodularity. Therefore, we can remove this element from $R$ immediately and do not need to consider this element any more for the sake of efficiency. And this is stated in line 16-18 of Algorithm  \ref{alg: Sample Threshold}.

\begin{algorithm}[H]
\caption{SDTGA}
\textbf{Input:} $f:2^\mathcal{N}\rightarrow\mathbb{R}_{\geq0}, \mathcal{N}, \mathcal{I}, r, p,  \epsilon$   \\
\textbf{Output:} A set $S\in\mathcal{I}$

\label{alg: Sample Threshold}
%\begin{multicols}{2}
\begin{algorithmic}[1]
\State $S \leftarrow \emptyset$, $R \leftarrow \emptyset$
\For{$u \in \mathcal{N}$}
	\State with probability $p$,
	\State $R \leftarrow R \cup \{u\}$
\EndFor
\State $d \leftarrow \max_{u \in R}f(u)$
\For {($ \theta = d; \theta \geq \frac{\epsilon}{r}d; \theta \leftarrow \theta(1-\epsilon))$}
	\For {$u \in R$}
		\If {$S\cup\{u\} \notin \mathcal{I}$}
			\State $R \leftarrow R \backslash \{u\}$
		\Else
			\If {$\Delta f(u|S) \geq \theta$}
				\State $S \leftarrow S \cup \{u\}$
				\State $R \leftarrow R \backslash \{u\}$
			\Else
				\If{$\Delta f(u|S) < \frac{\epsilon}{r}d$}
					\State $R \leftarrow R \backslash \{u\}$
				\EndIf	
			\EndIf
		\EndIf
	\EndFor
\EndFor \\
\Return $S$
\end{algorithmic}
%\end{multicols}
\end{algorithm}

	In order to better analyze Algorithm \ref{alg: Sample Threshold}, we make some modification to transform our algorithm into an equivalent version i.e. Algorithm \ref{alg: Equivalent Sample Threshold}.

\begin{algorithm}[H]
\caption{Equivalent SDTGA}
\textbf{Input:} $f:2^\mathcal{N}\rightarrow\mathbb{R}_{\geq0}, \mathcal{N}, \mathcal{I}, r, p,  \epsilon$   \\
\textbf{Output:} A set $S\in\mathcal{I}$

\label{alg: Equivalent Sample Threshold}
\begin{multicols}{2}
\begin{algorithmic}[1]
\State $S \leftarrow \emptyset$, $\mathcal{N}_s \leftarrow \emptyset$, $R \leftarrow \mathcal{N}$,\\ $C \leftarrow \emptyset$, $Q \leftarrow OPT$
\For {$u \in R$}
	\State with probability $p$,
	\State  $\mathcal{N}_s \leftarrow \mathcal{N}_s \cup \{u\}$
\EndFor
\State $d \leftarrow \max \limits_{u \in \mathcal{N}_s}f(u)$
\For {($ \theta = d; \theta \geq \frac{\epsilon}{r}d; \theta \leftarrow \theta(1-\epsilon))$}
	\For {$u \in R$}
		\If {$S\cup\{u\} \notin \mathcal{I}$}
			\State $R \leftarrow R \backslash \{u\}$
		\Else
			\If {$\Delta f(u|S) \geq \theta$}
				\State $c \leftarrow u$%, $S_c \leftarrow S$,  $C \leftarrow C \cup \{c\}$
				\State $S_c \leftarrow S$
				\State $C \leftarrow C \cup \{c\}$
				\If{$u \in \mathcal{N}_s$}
					\State $S \leftarrow S \cup \{c\}$
					\State $Q \leftarrow Q \cup \{c\}$
					\State Let $K_c\subseteq Q \backslash S$ be the smallest set such that $Q \backslash K_c \in \mathcal{I}$
				\Else
					\If{$c \in Q$}
						\State $K_c \leftarrow \{c\}$
					\Else
						\State $K_c \leftarrow \emptyset$
					\EndIf
				\EndIf
				\State $Q \leftarrow Q \backslash K_c$
				\State $R \leftarrow R \backslash \{c\}$
			\Else
				\If{$\Delta f(u|S) < \frac{\epsilon}{r}d$}
					\State $R \leftarrow R \backslash \{u\}$
				\EndIf	
			\EndIf
		\EndIf
	\EndFor
\EndFor \\
\Return $S$
\end{algorithmic}
\end{multicols}
\end{algorithm}

In Algorithm \ref{alg: Equivalent Sample Threshold}, the four variables $C$, $S_c$, $Q$ and $K_c$ are introduced just for the convenience of analysis and have no effect on the final output $S$. Therefore, Algorithm \ref{alg: Equivalent Sample Threshold} and Algorithm \ref{alg: Sample Threshold} are equivalent in fact. The rules of these variables are as following.
    
    $C$ is a set that contains all considered elements that have marginal values greater or equal to the current threshold $\theta$ in a certain iteration of Algorithm \ref{alg: Equivalent Sample Threshold} no matter whether they are added into $S$ or not. 
    
    $S_c$ is a set that contains the selected elements at the beginning of the current iteration. At the end of this iteration, $S_c = S \backslash \{c\}$ if $c$ is added into $S$ and $Q$, otherwise $S_c = S$.
    
    $Q$ is a set that bridges the relationship between the solution $S$ and the optimal solution $OPT$.
$Q$ starts as $OPT$ at the beginning of the algorithm and changes over time. Every element that is added into $S$ is also added into $Q$. At the same time, a set $K_c$ is removed from $Q$ at each iteration in order to keep the independence of $Q$ if an element $c$ is added into $Q$. Notice that, if an element $c$ is already in $Q$ and is considered but not added into $S$ at the current iteration, then this element $c$ should be removed from $Q$.

    $K_c$ is a set that is introduced to keep $Q$ independent. According to the property of $k$-extendible systems, the  Algorithm \ref{alg: Equivalent Sample Threshold} is able to remove a set $K_c \subseteq Q\backslash S$ which contains at most $k$ elements from $Q$ if an element is added into the currently independent set $Q$.

\subsection{Analysis}

% Theorem
\newtheorem{theorem}{Theorem}
\begin{theorem}
\label{thm:main_theorem}
The SDTGA achieves an approximation ratio of $p-\epsilon$ for monotone submodular maximization subject to $k$-extendible system constraints and of $p(1-p)-\epsilon$ for non-monotone cases using time complexity of $O(\frac{pn}{\epsilon}\ln\frac{r}{\epsilon})$, where $r$ is the largest size of the feasible solutions, $0<p \leq \frac{1}{1+k}$ is the sampling probability and $0 < \epsilon < p$.
\end{theorem}

% Time Complexity
	We start the analysis from the time complexity. Assume that there are totally $x$ number of iterations. Thus,
\begin{equation*}
\label{eq4}
(1-\epsilon)^x = \frac{\epsilon}{r} .
\end{equation*}
Solving the above equation, we get
\begin{equation*}
\label{eq4}
x = \frac{\ln\frac{r}{\epsilon}}{\ln\frac{1}{1-\epsilon}} \leq \frac{1}{\epsilon}\ln\frac{r}{\epsilon} .
\end{equation*}
And there are expectantly at most $p \cdot n$ submodular function evaluations in each iteration. Therefore, the time complexity of the algorithm is $O(\frac{pn}{\epsilon}\ln\frac{r}{\epsilon})$.

	In the following, we analyze the approximation guarantees of our algorithm in both monotone and non-monotone cases.

% Lemma 1
\newtheorem{lemma}{Lemma}

\begin{lemma}
\label{thm:lemma_1}
$\mathbb{E}[|K_c \backslash S|] \leq Pr_{max}$ where $Pr_{max} = \max(pk, 1-p)$.
\end{lemma}
\begin{IEEEproof}
There are two cases to analyze in terms of whether the current element $c$ is already in $Q$ at the beginning of the iteration in which Algorithm \ref{alg: Equivalent Sample Threshold} is considering $c$.

    1) $c \in Q$ at the beginning of the iteration. Then $K_c=\emptyset$ if $c$ is added to $S$ and $K_c=\{c\}$ if $c$ is not added to $S$. As $c$ is added to $S$ with probability $p$ and by the law of total probability, we get
\begin{equation*}
\mathbb{E}[|K_c \backslash S|] \leq p \cdot |\emptyset|+(1-p)|\{c\}|=1-p .
\end{equation*}

    2) $c \notin Q$ at the beginning of the iteration. Then $K_c$ contains at most $k$ elements if $c$ is added to $S$ because of the property of $k$-extendible systems. And if $c$ is not added to $S$ then $K_c=\emptyset$. We have
\begin{equation*}
\mathbb{E}[|K_c \backslash S|] \leq p \cdot k+(1-p)|\emptyset|=pk .
\end{equation*}

    In summary, $\mathbb{E}[|K_c \backslash S|] \leq \max(pk, 1-p)$.    
\end{IEEEproof}

% Lemma 2
\begin{lemma}
\label{thm:lemma_2}
$\mathbb{E}[f(S)] = p \sum \limits_{c \in C}\Delta f(c|S_c)$ .
\end{lemma}

\begin{IEEEproof}
According to the order by which the elements are added into $S$ and since $f$ is normalized i.e. $f(\emptyset) = 0$, $f(S)$ can be written as
\begin{equation*}
f(S) = f(\emptyset) + \sum \limits_{c \in S}\Delta f(c|S_c) = \sum \limits_{c \in S}\Delta f(c|S_c) .
\end{equation*}

    In each iteration, the element $c$ that is being considered is added into $S$ with probability $p$. If this element $c$ is added into $S$, then its marginal value $\Delta f(c|S_c)$ will be added to the current function value $f(S_c)$. Otherwise, the contribution of $c$ is 0. Therefore,
\begin{equation*}
f(S) = \sum \limits_{c \in C}[\Delta f(c|S_c)_{c \in S} + 0 \cdot \Delta f(c|S_c)_{c \in C\backslash S}] .
\end{equation*}
By the law of total probability, we have
\begin{equation*}
\mathbb{E}[f(S)] = p \cdot \sum \limits_{c \in C}\Delta f(c|S_c) + (1-p) \cdot 0 = p \cdot \sum \limits_{c \in C}\Delta f(c|S_c) .
\end{equation*}    
\end{IEEEproof}

% Lemma 3
\begin{lemma}
\label{thm:lemma_3}
$\mathbb{E}[f(S)] > \frac{(1-\epsilon)p}{(1-\epsilon^2)p+Pr_{max}}  \mathbb{E}[f(S \cup OPT)]$.
\end{lemma}

\begin{IEEEproof}
From Algorithm \ref{alg: Equivalent Sample Threshold} we can see that the set $Q$ is independent i.e. $Q\in \mathcal{I}$ and $S$ is a subset of $Q$ i.e. $S \subseteq Q$.  Therefore, we have $S \cup \{q\} \in \mathcal{I} ~ \forall q \in Q \backslash S$ by the property of independent systems and $|Q \backslash S| \leq r$. At the termination of Algorithm \ref{alg: Equivalent Sample Threshold}, $\Delta f(q|S) < \frac{\epsilon}{r}d ~ \forall q \in Q \backslash S$ and $f(S) \geq d$. Thus,
\begin{equation*}
\sum\limits_{q \in Q \backslash S}\Delta f(q|S) < \sum\limits_{q \in Q \backslash S}\frac{\epsilon}{r}d \leq \epsilon \cdot \frac{|Q \backslash S|}{r} f(S) \leq \epsilon \cdot f(S).
\end{equation*}
Let $Q \backslash S = \{q_1, q_2, \cdots, q_{|Q \backslash S|}\}$, then
\begin{align*}
f(S) &= f(Q) - \sum \limits_{i=1}^{|Q \backslash S|}\Delta f(q_i|S \cup \{q_1, \cdots, q_{i-1}\}) &\\
& \geq f(Q) - \sum\limits_{i=1}^{|Q \backslash S|}\Delta f(q_i|S)  &\text{(submodularity)} \\
& = f(Q) - \sum\limits_{q \in Q \backslash S}\Delta f(q|S) \\
& > f(Q) - \epsilon \cdot f(S).
\end{align*}
Therefore, $f(S) > \frac{1}{1+\epsilon}f(Q)$.

    In some iteration and given the current threshold $\theta$, if $c$ is being considered then it implies that
\begin{equation}
\label{threshold greater} 
\Delta f(c|S_c) \geq \theta .
\end{equation}
While if an element  $q \in K_c \backslash S$ was not selected before this iteration, then
\begin{equation}
\label{threshold less} 
\Delta f(q|S_c) < \theta / (1-\epsilon) .
\end{equation}
Combining (\ref{threshold greater}) and (\ref{threshold less}) we have
\begin{equation} 
\label{c vs o}   
\Delta f(c|S_c) > (1-\epsilon) \Delta f(q| S_c) ~ \forall q \in K_c \backslash S.
\end{equation}

    Additionally, any element can be removed from $Q$ at most once. In other words, the element that is contained in $K_c$ at one iteration is always different from other iterations when $K_c$ is not empty. Therefore, the sets $\{K_c\}_{c \in C}$ and sets $\{K_c \backslash S\}_{c \in C}$ are disjoint. And by definition of $Q$, we can rewrite $Q$ as
\begin{equation}
\label{definition of O}
Q = (OPT \backslash \cup_{c \in C}K_c) \cup S = (S \cup OPT) \backslash \cup_{c \in C}(K_c \backslash S).
\end{equation}
Denote $C$ as $\{c_1, c_2, \cdots, c_{|C|}\}$. Then, it holds that $S_{c_i} \subseteq S \subseteq (S \cup OPT) \backslash \cup_{c \in C}(K_c \backslash S)$. Using equation (\ref{definition of O}), we have

\begin{align*}
f(Q) &= f((S \cup OPT) \backslash \cup_{c \in C}(K_c \backslash S)) &\\
&= f(S \cup OPT)- \Delta f(\cup_{c \in C}(K_c \backslash S)|(S \cup OPT)\backslash \cup_{c \in C} (K_c \backslash S)) &\\
&= f(S \cup OPT)- \sum \limits_{i=1}^{|C|} \Delta f((K_{c_i} \backslash S)|(S \cup OPT)\backslash \cup_{1\leq j \leq i} (K_{c_j} \backslash S)) &\\
& \geq  f(S \cup OPT) - \sum \limits_{i=1}^{|C|} \Delta f((K_{c_i} \backslash S)|S_{c_i})  &\text{(submodularity)} \\
& \geq  f(S \cup OPT) - \sum \limits_{i=1}^{|C|} \sum \limits_{q \in K_{c_i} \backslash S}\Delta f(q|S_{c_i})  &\text{(submodularity)} \\
& = f(S \cup OPT) - \sum \limits_{c \in C} \sum \limits_{q \in K_c \backslash S}\Delta f(q| S_c) &\\
& > f(S \cup OPT) - \sum \limits_{c \in C} \sum \limits_{q \in K_c \backslash S}\frac{1}{1-\epsilon}\Delta f(c| S_c) &\text{(inequation (\ref{c vs o}))} \\
& = f(S \cup OPT) - \sum \limits_{c \in C}|K_c \backslash S|\frac{1}{1-\epsilon}\Delta f(c| S_c) .
\end{align*}
    
    By taking expectation over $f(S)$, we obtain
\begin{align*}
\mathbb{E}[f(S)] &> \frac{1}{1+\epsilon} \mathbb{E}[f(Q)] & \\
&> \frac{1}{1+\epsilon} \mathbb{E}[f(S \cup OPT) - \sum \limits_{c \in C}|K_c \backslash S|\frac{1}{1-\epsilon}\Delta f(c|S_c)] &\\
&= \frac{1}{1+\epsilon} \mathbb{E}[f(S \cup OPT)] - \frac{1}{(1+\epsilon)(1-\epsilon)} \mathbb{E}[|K_c \backslash S|] \sum\limits_{c \in C}\Delta f(c| S_c) &\\
&\geq \frac{1}{1+\epsilon} \mathbb{E}[f(S \cup OPT)] - \frac{1}{(1+\epsilon)(1-\epsilon)} \cdot Pr_{max}\cdot \sum\limits_{c \in C}\Delta f(c| S_c) &\text{(Lemma \ref{thm:lemma_1})} \\
&= \frac{1}{1+\epsilon} \mathbb{E}[f(S \cup OPT)] - \frac{1}{(1+\epsilon)(1-\epsilon)} \cdot \frac{ Pr_{max}}{p} \cdot \mathbb{E}[f(S)]. &\text{(Lemma \ref{thm:lemma_2})}
\end{align*}
The result is clear by rearranging the above inequation.    
\end{IEEEproof}

    We are now ready to finish the proof of Theorem \ref{thm:main_theorem}.
\begin{IEEEproof} In order to get the approximation guarantees for both monotone and non-monotone submodular objective functions, we need to analyze the relationship between $f(S \cup OPT)$ and $f(OPT)$ respectively. If $f$ is monotone, then 
\begin{equation*}
f(S \cup OPT) \geq f(OPT).
\end{equation*}
Otherwise, define a new submodular and non-monotone function $h:2^\mathcal{N}\rightarrow\mathbb{R}_{\geq0}$ as $h(X)=f(X\cup OPT) ~\forall X \subseteq \mathcal{N}$. Since $S$ contains every element with probability at most $p$ and according to Claim \ref{thm:main_claim}, we get
\begin{equation*}
\mathbb{E}[f(S \cup OPT)]=\mathbb{E}[h(S)] \geq (1-p)h(\emptyset)=(1-p)f(OPT).
\end{equation*}

    Recall that, $Pr_{max}= \max(pk, 1-p)$. Thus when $\frac{1}{1+k}<p<1$, $Pr_{max}=pk$ and we have
\begin{align*}
\mathbb{E}[f(S)] &> \frac{(1-\epsilon)p}{(1-\epsilon^2)p+Pr_{max}} \cdot \mathbb{E}[f(S \cup OPT)] &\text{(Lemma \ref{thm:lemma_3})} \\
& = \frac{1-\epsilon}{1+k-\epsilon^2}\cdot\mathbb{E}[f(S \cup OPT)] \\
& > (\frac{1}{1+k}-\epsilon)\cdot\mathbb{E}[f(S \cup OPT)].
\end{align*}
The expected approximation ratios are
\begin{equation*}
\mathbb{E}[f(S)] > \left \{ \begin{array}{lcl} (\frac{1}{1+k}-\epsilon)f(OPT) & \mbox{if} & f ~ \mbox{is monotone}  \\ (\frac{1}{1+k}-\epsilon)(1-p)f(OPT) & \mbox{if} & f ~ \mbox{is non-monotone}.\end{array}\right.
\end{equation*}
While $0< p \leq \frac{1}{1+k}$, $Pr_{max}=1-p$ we have
\begin{align*}
\mathbb{E}[f(S)] &> \frac{(1-\epsilon)p}{(1-\epsilon^2)p+Pr_{max}} \cdot \mathbb{E}[f(S \cup OPT)] &\text{(Lemma \ref{thm:lemma_3})} \\
& = \frac{(1-\epsilon)p}{(1-\epsilon^2)p+1-p} \cdot \mathbb{E}[f(S \cup OPT)] \\
& > (p-\epsilon) \cdot \mathbb{E}[f(S \cup OPT)].
\end{align*}
The expected approximation ratios are
\begin{equation*}
\mathbb{E}[f(S)] > \left \{ \begin{array}{lcl} (p-\epsilon) \cdot f(OPT) & \mbox{if} & f ~ \mbox{is monotone} \\ {[p(1-p)-\epsilon]\cdot f(OPT)} & \mbox{if} & f ~ \mbox{is non-monotone}. \end{array}\right.
\end{equation*}

	We can see from the results that there is no advantage if $\frac{1}{1+k}<p<1$ because as $p$ increases the computational complexity increases while the approximation ratio for non-monotone case decreases. When $0<p\leq\frac{1}{1+k}$, there is some trade-off between the approximation ratio and computational complexity. Therefore, we just abandon the case of $p>\frac{1}{1+k}$. While $p=\frac{1}{1+k}$, the proposed algorithm achieves the best approximation ratio for both monotone ($\frac{1}{1+k}-\epsilon$) and non-monotone ($\frac{k}{(1+k)^2}-\epsilon$) submodular objective functions with time complexity of $O(\frac{n}{(1+k)\epsilon}\ln\frac{r}{\epsilon})$.

\end{IEEEproof}  

%====================================================================%
%Experimental Results
\section{Conclusions}

	In this paper, we have presented a very fast algorithm (SDTGA) for maximizing general non-negative submodular functions subject to $k$-extendible system constraints. Our algorithm achieves an expected approximation ratio of ($\frac{k}{(1+k)^2}-\epsilon$) for general non-monotone submodular objective functions with only $O(\frac{n}{(1+k)\epsilon}\ln\frac{r}{\epsilon})$ value oracle calls which is currently the most efficient in theory. We believe our algorithm has made an important progress towards discrete optimization problems where the sizes of data sets are enormous such as the applications of machine learning and big data science.

%====================================================================%
% References

\end{document}